\documentstyle[aps,preprint]{revtex}
\begin{document}
\draft
\preprint{\hspace*{\fill} RESCEU-95-11}
\title{Sensitive germanium thermistors for cryogenic thermal
detector of Tokyo dark matter search programme}
\author{Wataru Ootani, Yutaka Ito, Keiji Nishigaki, Yasuhiro
Kishimoto, and Makoto Minowa}
\address{Department of Physics, Graduate School of Science, University of
Tokyo, 7-3-1 Hongo, Bunkyo-ku, Tokyo 113, Japan}
\author{Youiti Ootuka}
\address{Cryogenic Center, University of Tokyo,
2-11-16 Yayoi, Bunkyo-ku, Tokyo 113, Japan}
\maketitle
\begin{abstract}
Sensitive n-type and p-type germanium thermistors
were fabricated by the melt doping technique and by the neutron
transmutation doping (NTD) technique, respectively, aiming at a use for the
cryogenic
thermal detector, or bolometer of Tokyo dark matter search programme.
We report on the measurements of the sensitivities of
these thermistors.
In particular, the p-type thermistors are sensitive enough to scale up
our existing prototype LiF bolometer and realize a
multiple array of the bolometers with the total absorber mass of about
1\,kg.
\end{abstract}

\section{Introduction}
\label{sec:intro}
One of the most important issues
of the modern physics is the nature of the dark matter.
Weakly interacting massive particles (WIMPs) is considered as the
leading candidates for the particle dark matter.
A direct search for WIMPs using a cryogenic thermal detector, or
bolometer is planned.
This programme is now under way and a prototype of the detector
is already completed\cite{LiF bolometer}.
A bolometer consists of
an absorber in which
a WIMP scatters off a nucleus elastically and a thermistor which senses
the resulting temperature rise of the absorber.
Operation at very low temperature leads to small specific heat of the
absorber.
A bolometer offers
significant advantages for the dark matter search because of its high energy
resolution, low energy threshold, and flexibility on a choice of the
absorber materials. \\

We developed a prototype bolometer with a 2.8-g LiF
absorber\cite{LiF bolometer} on
the grounds that an elastic scattering cross section of the axially-coupled
WIMPs on $^{19}$F (natural abundance 100\%) is
fairly larger than on other nuclei like $^{73}$Ge, $^{23}$Na, and
$^{27}$Al\cite{LiF bolometer,Ellis}.
Its energy
resolution was 3.8\,keV (rms) for 60-keV $\gamma$
rays. Absorber mass of about 1\,kg is, however, required for searching
for the WIMPs. We are planning to scale up the prototype bolometer and
construct a multiple array of the bolometers with the total mass of
about 1\,kg. In realizing this scaling-up of the bolometer, we need more
sensitive thermistors, as well as a reduction of the noise level. \\

To our knowledge, there is no ready-made thermistor that can be used
below 50mK with a sufficient sensitivity and has a small size which
leads to small specific heat.
Many attempts to develop sensitive thermistors  have,
therefore, been made so far by different groups, using
superconductors\cite{edge}, heavily doped
semiconductors\cite{lblntd}, or thin alloy films\cite{thin alloy
films}. A heavily doped
germanium thermistor is used in our prototype LiF bolometer\cite{LiF
bolometer}.\\

Electrical conductance in heavily doped and compensated semiconductors
at low temperature is dominated by variable hopping
conductance\cite{Shklovskii}. In this conduction regime, charge
carriers tunnel from an occupied dopant site to an unoccupied
dopant site. The uniformity of the dopant distribution and the ratio
of the minority dopants to the majority dopants (compensation ratio)
are critical
parameters as well as the dopant concentration. The temperature dependence of
the resistance $R$ of doped germanium predicted in this conduction regime is
\begin{equation}
R\left(T\right)=R_0\exp\left(\frac{\Delta}{T}\right)^{\frac{1}{2}},
\label{eq:rt}
\end{equation}
where $R_0$ and $\Delta$ are constant parameters, and $T$ is the
temperature in Kelvin.
The dimensionless sensitivity of a thermistor can be defined by the
logarithmic slope $n$:
\begin{equation}
n=-\frac{d\left(\ln R\right)}{d\left(\ln T\right)}
=\frac{1}{2}\left(\frac{\Delta}{T}\right)^{\frac{1}{2}}.
\label{eq:sensitivity}
\end{equation}\\

A fabrication of thermistors by doping semiconductors is quite
delicate
because an impurity concentration very close to the metal-insulator
transition is required. It also needs to be uniform everywhere in a
sample. Furthermore, an amount of thermistors
with similar characteristics are needed for the construction of
a multiple array of the bolometers. \\

The present paper reports on the measurements of the sensitivities of
the germanium thermistors fabricated by the melt doping technique and
the neutron transmutation doping (NTD)
technique. Particularly, the latter were very sensitive and
reproducible as we shall see later in Sec.\ \ref{sec:measurement}.
The thermistors by the NTD technique have been developed so far
by LBL group\cite{lblntd}, but ours have slightly higher
sensitivities.

\section{Manufacture of n-type germanium thermistors by melt doping technique}
\label{sec:melt}
The melt doping is one of the conventional techniques to dope
semiconductors. Our n-type germanium thermistors which is
used in our prototype LiF bolometer are doped with
antimony (n-type dopant) and compensated with copper (p-type dopant).\\

The starting materials available to us were germanium wafers
doped with antimony,
whose resistivities ranged from $2.55\times10^{-2}$ to
$3.22\times10^{-2}$\,$\Omega\, \mbox{cm}$
at room temperature.
This resistivity range corresponds to the charge carrier concentration from
$1.26\times10^{17}$ to $1.00\times10^{17}$\,cm$^{-3}$\cite{rho vs n},
which is far from the metal-insulator transition\cite{transition}.
In order to reduce the effective charge carrier concentration
the samples were compensated with copper which is
opposite type dopant.\\

The thermistors were manufactured by the following procedure.
A chip with dimensions of $1.5\times 1.5\times 1^t$\,mm$^3$ was cut
from the wafer. A 1\% aqueous solution of copper nitrate was smeared
on one square face of the
chip. The chip was annealed at 800\,$^{\circ}$C for twenty minutes
to diffuse copper. A solid solubility limit of copper in germanium at
800\,$^{\circ}$C fixes the copper concentration of the chip at
$1.9\times10^{16}$\, cm$^{-3}$\cite{solubility}.
The resulting charge carrier concentration and the
compensation ratio of each sample are listed in Table \ref{table:melt}. \\

In order to form ohmic contacts
antimony pads with a thickness of 1000\,\AA\ were evaporated on two areas on
one square
face and gold pads with a thickness of 1000\,\AA\ followed,
as shown in Fig.\ \ref{fig:thermistor}. The chip was annealed
at 600\,$^{\circ}$C for thirty minutes to activate the contacts.
Gold wires with a diameter of 50\,$\mu$m were attached
to the contacts by the spot welding.

\section{Manufacture of p-type germanium thermistors by NTD technique}
\label{sec:ntd}
The neutron transmutation doping (NTD) which is a technique to dope
semiconductors by partial transmutation of its stable
isotopes to dopants via thermal
neutron capture, was first put into practical use in high power device
with silicon\cite{NTD}. Since the NTD technique provides uniform dopants
concentration in the crystal, large area silicon device became
available. \\

The NTD process can provide much more uniform distribution of the dopants in
germanium.
Impurity conduction in the NTD germanium was experimentally studied
by Fritzsche {\it et al}\cite{Fritzsche}.
Thermal neutrons from a nuclear reactor are captured by the
five isotopes contained in natural
germanium;$^{70}$Ge,$^{72}$Ge,$^{73}$Ge,$^{74}$Ge, and
$^{76}$Ge. The available
information of these reactions are listed in
Table \ref{table:NTD reaction}\cite{table of isotopes},
which indicates $^{70}$Ge,$^{74}$Ge, and $^{76}$Ge are transmuted
to the p-type majority dopants or n-type minority dopants
and the others are not.
Net hole concentration can be controlled by the amount of the irradiated
thermal neutrons. Furthermore, the moderate cross sections
of thermal-neutron capture allow one to dope a large volume crystal of
germanium with
extremely uniform dopant concentration. The compensation ratio is
fixed to 0.32 due to the fixed abundance of germanium isotopes and
the cross sections of thermal-neutron capture.\\

Pure germanium single crystals with resistivities of
over 50\,$\Omega\,$cm were exposed to the thermal neutrons in
the JRR-3M reactor at Japan Atomic Energy Research Institute.
The amount of the irradiated thermal
neutrons and the resulting net hole concentration are listed in
Table \ref{table:NTD}.\\

After the exposure a chip with a dimension of $1.5\times 1\times 1^t$\,mm$^3$
was cut from the crystal. In order to remove the undesired
radiation damage due to fast neutrons, the chip was annealed
at 400\,$^\circ$C for ninety n[minutes. Boron (p-type dopant) ions with a
dose
of $3\times10^{14}$\,cm$^{-2}$ were implanted on two areas on one face of
the chip to a depth of 1200\,\AA\ in order to form
heavily doped regions which yield ohmic contacts. Gold pads as thick as
1500\,\AA\ were evaporated on the implanted sites.
The contacts were activated by annealing
the chip at 250\,$^\circ$C for ninety minutes.
Gold wires with a diameter of 50\,$\mu$m were attached
to the contacts in a similar way as the n-type thermistors
(Fig.\ \ref{fig:thermistor}).

\section{Measurements and results}
\label{sec:measurement}
The measurements of the temperature dependence of the resistance of
the thermistors were carried out in an Oxford Instruments refrigerator
equipped with a $^3$He/$^4$He dilution unit with a cooling power of
600\,$\mu$W at 100\,mK.
The whole system except the pumping system is housed in an
electoromagnetic shielding room.\\

The fabricated n-type or p-type thermistors were mounted onto a
copper plate which was thermally anchored to the mixing chamber of the
dilution unit, and cooled down to between 30 and 110\,mK.
The temperature of the mixing chamber was monitored
by a carbon resistor, a $^{60}$Co nuclear orientation thermometer, or
both.
The thermistor was glued with GE varnish to a Kapton foil
with a thickness of 7.5\,$\mu$m which was glued to the copper plate.
A Kapton foil is an insulating foil with a good thermal conductivity.\\

The thermistor was biased at a current between 50\,pA and 100\,nA with
load resistors by an external battery.
Since a nonlinearity of the current-voltage relations of the
thermistor was conspicuous at low temperature, the resistance was
determined by the zero bias resistance $\left(dV/dI\right)_{I=0}$.
The measured temperature dependence of the resistance ($R$-$T$curve)
of the n-type germanium
thermistors by the melt doping technique
is given in Fig.\ \ref{fig:tdy_rt} and of the p-type germanium
thermistors by the NTD
technique in Fig.\ \ref{fig:ntd_rt}.
The solid lines in the figures are the best
fits to the data. The parameters $R_0$ and $\Delta$ in Eq.\ (\ref{eq:rt})
and the dimensionless sensitivities $n$ at 20\,mK extracted from
these $R$-$T$curves are presented in Table \ref{table:melt}
for the n-type thermistors and Table \ref{table:NTD} for
the p-type thermistors. \\

Fig.\ \ref{fig:tdy_rt} shows that for the n-type thermistors
both the slope of the $R$-$T$curve and the resistance
at the same temperature decrease with the charge carrier
concentration. Therefore, one can see that the estimated charge carrier
concentration is a parameter to control the sensitivity of
the n-type germanium thermistor by the melt doping technique.
While the dimensionless sensitivity $n$ is proportional to
the slope of the $R$-$T$curve as Eq.\ (\ref{eq:sensitivity}) indicates,
the resistance from a few M$\Omega$ to a few hundred M$\Omega$ is
desirable at the operating temperature in order to avoid electronic noise such
as Johnson noise. From this point of view
the sample MDGe3 is most favorable among
the measured n-type thermistors and is used in our prototype LiF
bolometer.
The drawback in the manufacture of the n-type
thermistors by the melt doping technique is that the dopants
distribution it provides is not extremely uniform, and it leads to poor
reproducibility in the $R$-$T$ characteristics.
This poor reproducibility was observed in our experiments.\\

\begin{figure}
\caption{$R$-$T$ curves of five NTDGe3 thermistors with the same
irradiation dose of $3.42\times10^{18}$ neutrons cm$^{-2}$. Solid
lines are the best
fits to the data.}
\label{fig:reproducibility}
\end{figure}
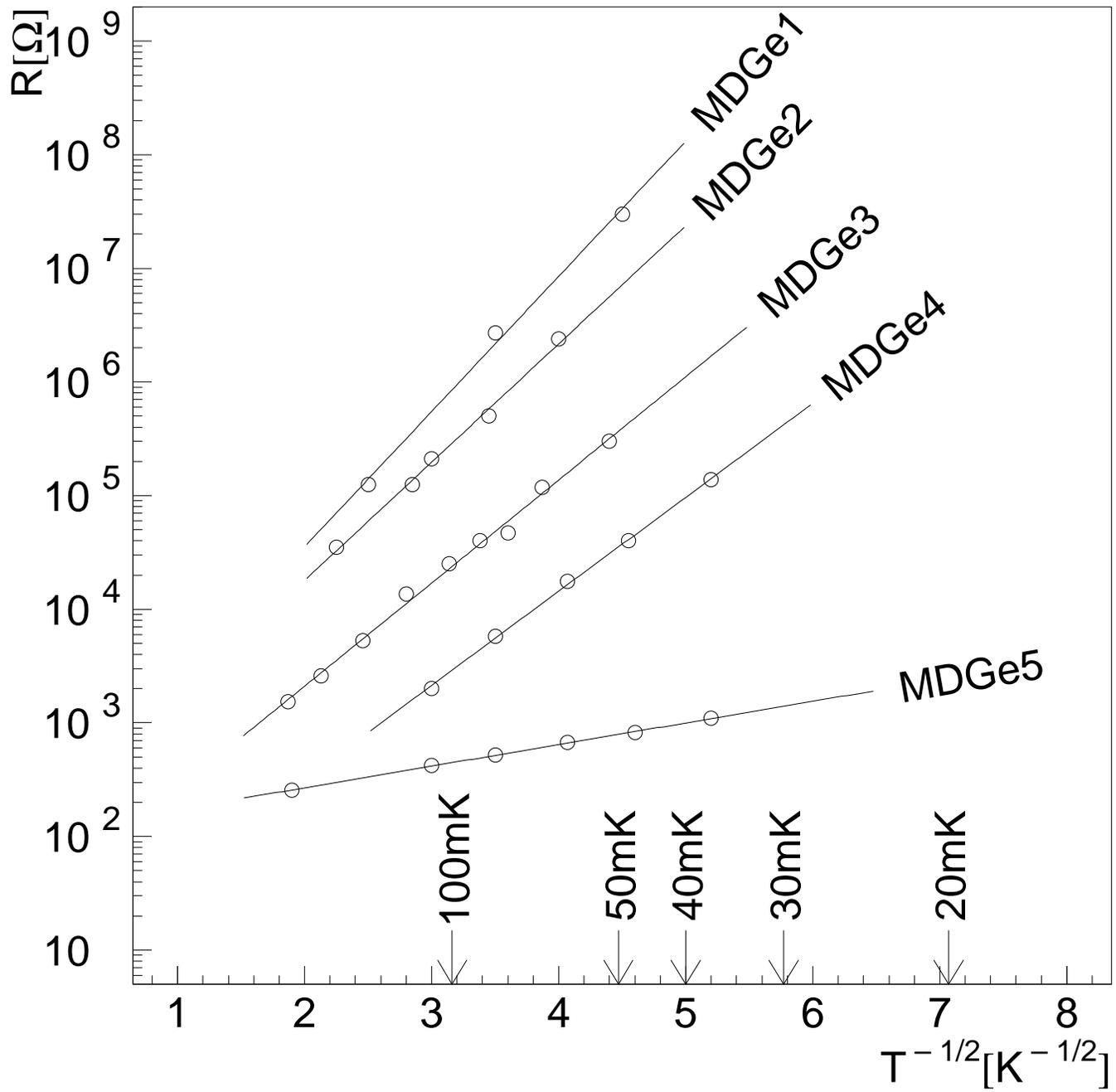
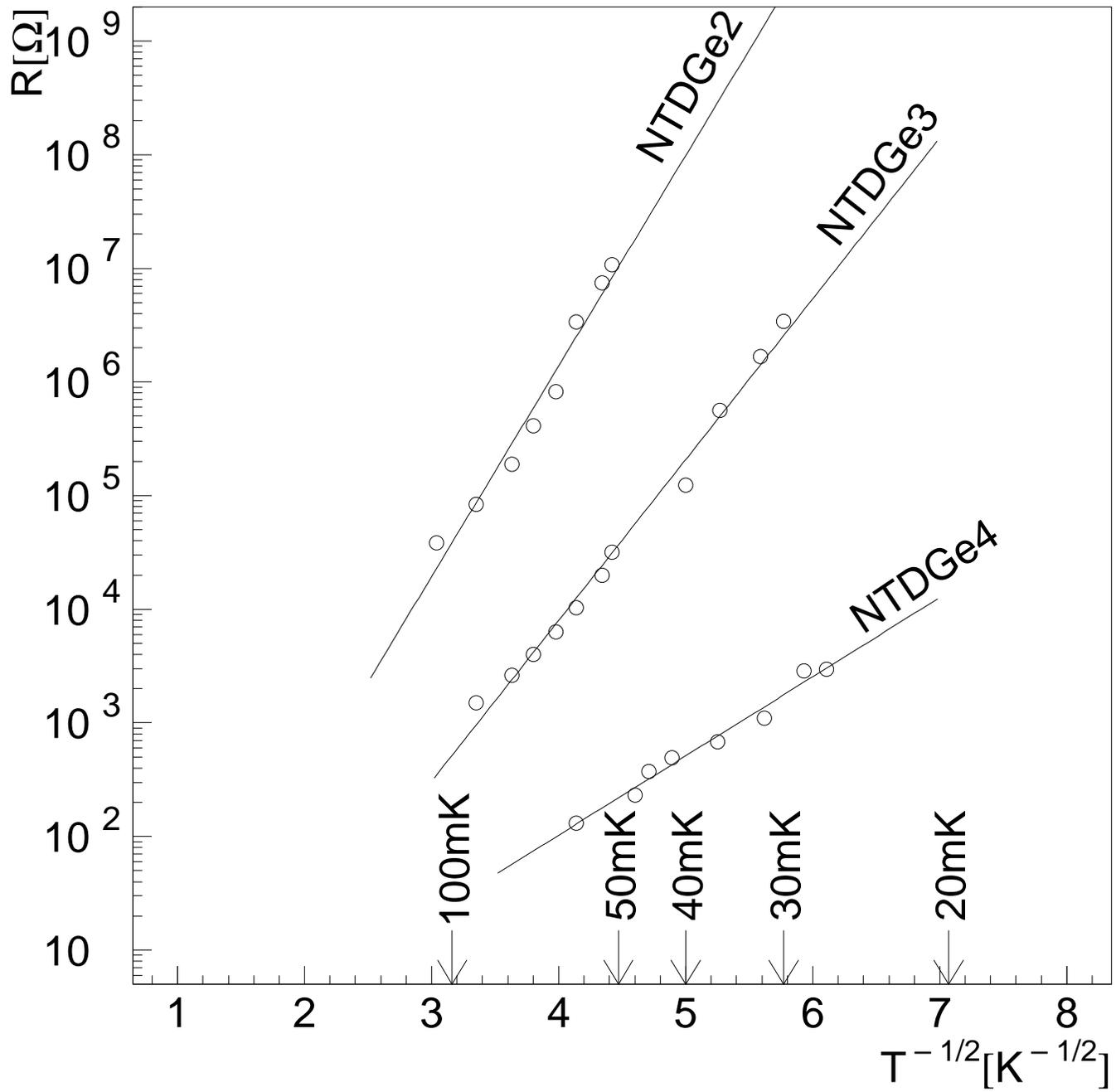
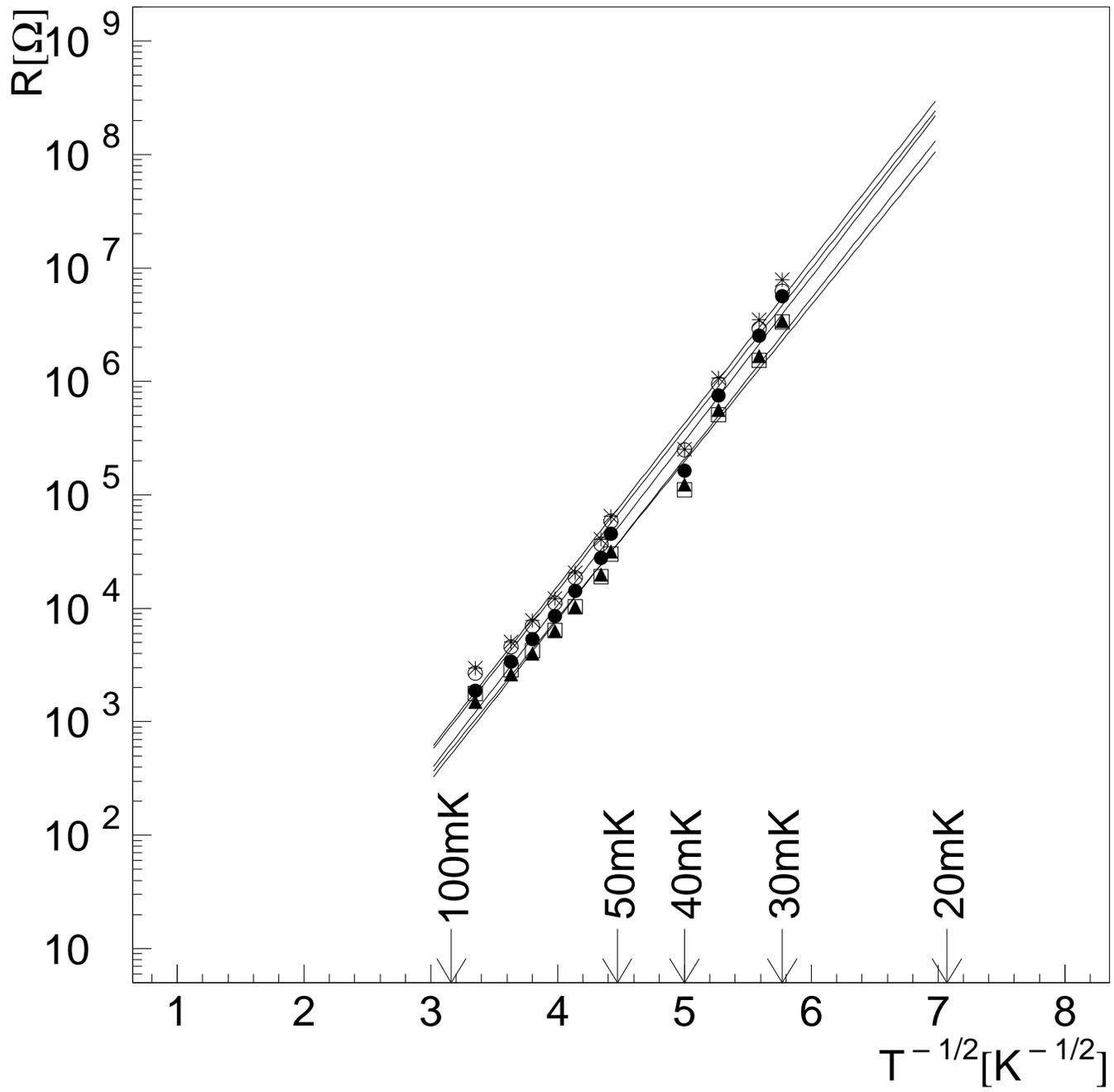
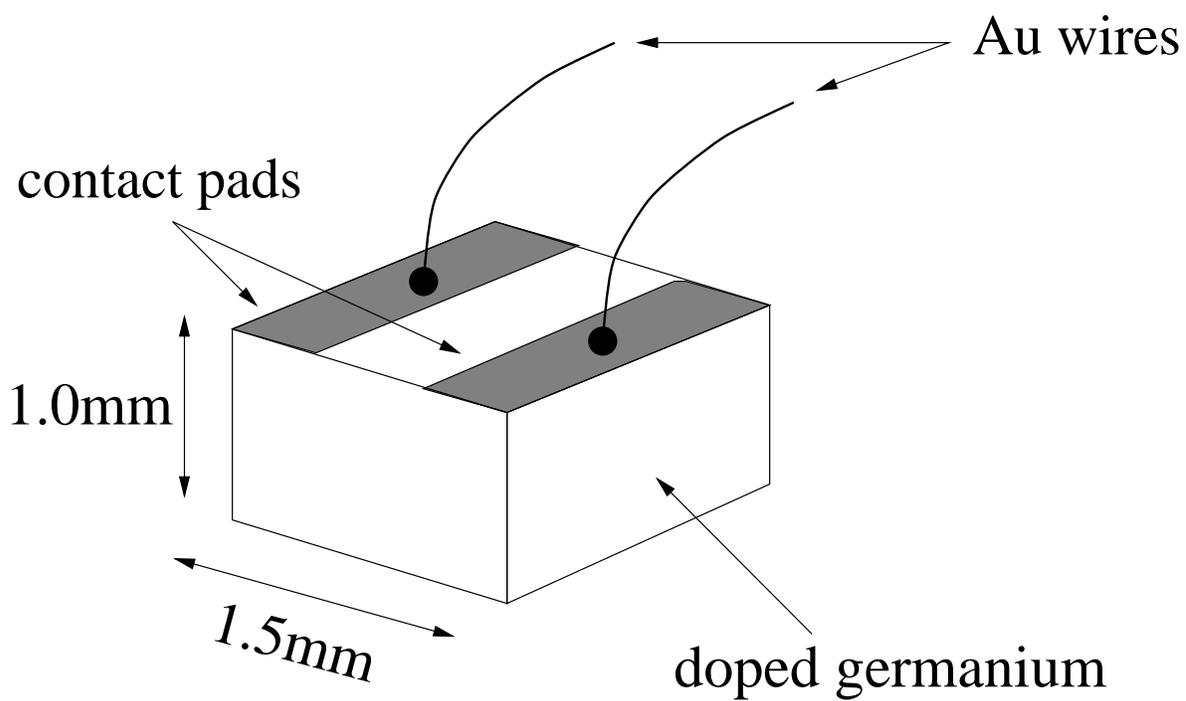

Fig.\ \ref{fig:ntd_rt} also shows that for the p-type thermistors both
the slope of the the $R$-$T$ curve and the value of
$R$ at the same temperature decrease with the charge carrier
concentration in the same way as the n-type thermistors.
{}From the noise consideration mentioned above, the sample NTDGe3
is most favorable. The sensitivity of the p-type thermistor was much
higher than that of the n-type thermistor.
Furthermore, the $R$-$T$ characteristics were very reproducible for the
thermistors exposed to the same amount of the thermal neutrons.
The $R$-$T$curves of five NTDGe3 thermistors with the same irradiation
dose of $3.42\times10^{18}$\,neutrons cm$^{-2}$ are shown in
Fig.\ \ref{fig:reproducibility}.
It is apparent that these five thermistors have similar $R$-$T$
characteristics.
This good reproducibility of the p-type thermistors would provide a
significant advantage in constructing a multiple array of the
bolometers.

\section{Summary and discussions}
\label{sec:summary and discussions}
In summary, we fabricated n-type and p-type germanium thermistors
by the melt doping technique and by the NTD technique, respectively, and
measured the $R$-$T$
characteristics of the thermistors.
We controlled the sensitivity of the n-type thermistor by
the measured resistivity at room temperature
before compensating and of the p-type thermistor
by the amount of the irradiated thermal neutrons.\\

The p-type thermistor had much higher sensitivity
and better reproducibility than the n-type thermistor.
These differences are considered to be primarily due to the facts that
the impurity distribution in the NTD is much more uniform than that in the
melt doping
and the compensation ratio of the p-type thermistor is different from that of
the n-type thermistor. The sensitivity of the NTDGe3 thermistor, which
has most suitable $R$-$T$ characteristics for our purpose,
is by a factor of 1.2 higher than that of
the NTD thermistor with similar resistance fabricated
by LBL group\cite{lblntd}.
The differences in the geometry of the thermistor and the
configuration of the contacts may cause this small difference in the
sensitivity.\\

The improvement on the resolution of the prototype 2.8-g LiF bolometer
in using the NTDGe3 thermistor can be estimated.
Assuming the same input power into the thermistor and the thermistor
temperature of 20\,mK, the resolution could be improved
by an order of magnitude. This improvement could allow us to realize
the absorber mass of 30\,g and construct multiple array of the
bolometers with similar sensitivities to perform the dark matter
search with a total absorber mass of about 1\,kg.

\section*{Acknowledgements}
We would like to thank Prof.\ Y. Ito and H. Sawahata for
irradiating germanium in the JRR-3M reactor at Japan Atomic Energy
Research Institute, Y. Koizumi for very useful discussions
on radioactive contaminations of the irradiated germanium, and
Dr.\  H. Bando for providing us with the dilution refrigerator.
This work has been supported by the Inter-University Program for the
Joint Use of JAERI Facilities.
This work is financially supported by the CASIO Science Promotion Foundation,
the Iwatani Naoji Foundation's Research Grant,
the Yamada Science Foundation, and
the Grant-in-Aid for Scientific Research (A)
and the Graint-in-Aid for Developmental Scientific Research by the
Japanese Ministry of Culture, Education and Science.

\begin{table}
\caption{Parameters of the n-type thermistors by the melt doping technique}
\begin{tabular}{cccccc}
Sample
&
\begin{tabular}{c} Charge carrier \\concentration \\(cm$^{-3}$)
\end{tabular}
&
\begin{tabular}{l} Compensation\\ratio
\end{tabular}
&
\begin{tabular}{c} $R_0$\\$(\Omega)$
\end{tabular}
&
\begin{tabular}{c} $\Delta$\\$(\mbox{K})$
\end{tabular}
&
\begin{tabular}{c} Sensitivity\\ $n$
\end{tabular}
\\ \hline
MDGe1 &$8.1\times10^{16}$&0.19&$148$&7.51&9.69\\\hline
MDGe2 &$8.9\times10^{16}$&0.18&$148$&5.74&8.47\\\hline
MDGe3 &$9.1\times10^{16}$&0.17&$32.1$&4.37&7.39\\\hline
MDGe4 &$1.1\times10^{17}$&0.15&$6.97$&3.64&6.75\\\hline
MDGe5 &$1.2\times10^{17}$&0.14&$112$&0.192&1.55\\
\end{tabular}
\label{table:melt}
\end{table}
\newpage

\begin{table}
\caption{Reactions of germanium isotopes in NTD process}
\begin{tabular}{ccclc}
Isotope&\begin{tabular}{c} Abundance\\(\%)\end{tabular}
&\begin{tabular}{c} Cross section\\(barn)\end{tabular}
&\multicolumn{1}{c}{Reaction}
& \begin{tabular}{c} Dopant\\type\end{tabular}\\\hline
$_{32}^{70}$Ge & 20.5 &3.25&$_{32}^{70}\mbox{Ge}
(n,\gamma)_{32}^{71}\mbox{Ge}^{11.2d}_{\longrightarrow}
\:  _{31}^{71}\;\mbox{Ga}$ & p \\ \hline
$_{32}^{72}$Ge & 27.4 & 1.0 &$_{32}^{72}\mbox{Ge}
(n,\gamma)_{32}^{73}\mbox{Ge}$ &$-$ \\ \hline
$_{32}^{73}$Ge & 7.8 & 15 &$_{32}^{73}\mbox{Ge}
(n,\gamma)_{32}^{74}\mbox{Ge}$ &$-$ \\ \hline
$_{32}^{74}$Ge &36.5 &0.52& $_{32}^{74}\mbox{Ge}
(n,\gamma)_{32}^{75}\mbox{Ge}
^{82.8m}_{\longrightarrow}\:  _{33}^{75}\;\mbox{As}$ & n \\ \hline
$_{32}^{76}$Ge &7.8 &0.16& $_{32}^{76}\mbox{Ge}
(n,\gamma)_{32}^{77}\mbox{Ge}
^{11.3h}_{\longrightarrow}\:  _{33}^{77}\;\mbox{As}
^{38.8h}_{\longrightarrow}\:  _{34}^{77}\;\mbox{Se}$ &  n\\
\end{tabular}
\label{table:NTD reaction}
\end{table}
\newpage

\begin{table}
\caption{Parameters of the p-type germanium thermistors by NTD technique}
\begin{tabular}{ccccccc}
Sample
&
\begin{tabular}{c} Irradiation \\ (neutrons cm$^{-2}$)
\end{tabular}
&
\begin{tabular}{c} Charge carrier \\concentration \\(cm$^{-3}$)
\end{tabular}
&
\begin{tabular}{l} Compensation\\ratio
\end{tabular}
&
\begin{tabular}{c} $R_0$\\$(\Omega)$
\end{tabular}
&
\begin{tabular}{c} $\Delta$\\$(\mbox{K})$
\end{tabular}
&
\begin{tabular}{c} Sensitivity\\ $n$
\end{tabular}
\\ \hline
NTDGe2&$2.85\times10^{18}$&$5.70\times10^{16}$&0.32
&$5.35\times10^{-2}$&18.2&15.1\\\hline
NTDGe3 &$3.42\times10^{18}$&$6.84\times10^{16}$&0.32
&$1.74\times10^{-2}$&10.6&11.5\\\hline
NTDGe4 &$4.11\times10^{18}$&$8.22\times10^{16}$&0.32
&$1.68\times10^{-1}$&2.58&5.68\\
\end{tabular}
\label{table:NTD}
\end{table}
\begin{figure}
\caption{Schematic drawing of the fabricated n-type or p-type
germanium thermistor.}
\label{fig:thermistor}
\end{figure}

\begin{figure}
\caption{Measured temperature dependence of the resistance of the n-type
germanium thermistors by the melt doping technique. Solid lines are
the best fits to the data.}
\label{fig:tdy_rt}
\end{figure}

\begin{figure}
\caption{Measured temperature dependence of the resistance of the p-type
germanium thermistors by the NTD technique. Solid lines are the best
fits to the data.}
\label{fig:ntd_rt}
\end{figure}

\end{document}